\documentstyle[11pt]{article}

\font\twelve=cmbx10 at 15pt
\font\ten=cmbx10 at 12 pt

\renewcommand{\thefootnote}{\fnsymbol{footnote}}

\begin{document}

\begin{titlepage}

\begin{center}

{\ten Centre de Physique Th\'eorique\footnote{Unit\'e Propre de
Recherche 7061} - CNRS - Luminy, Case 907}

{\ten F-13288 Marseille Cedex 9 - France }

\vspace{2 cm}

{\twelve BARYON MASSES FROM QCD CURRENT CORRELATORS AT {\large $T\neq 0$}}

\vspace{0.3 cm}
\setcounter{footnote}{0}
\renewcommand{\thefootnote}{\arabic{footnote}}

{\bf V. L. ELETSKY\footnote{Institute of Theoretical and Experimental
Physics, B. Cheremushkinskaya 25, \\117259 Moscow, Russia}}

\vspace{2 cm}

{\bf Abstract}

\end{center}

Correlation functions of QCD currents with quantum numbers
of nucleon and $\Delta$-isobar are
considered at finite temperatures.
Corrections of order $T^4$ to the correlators are calculated and
interpreted in terms of thermal mass shifts using a QCD sum rules
type of argument. In both cases the masses decrease with $T$.

\vspace{1.5 cm}


\bigskip

\noindent Number of figures : 2

\bigskip

\noindent December 1994

\noindent CPT-94/P.3139

\bigskip

\noindent anonymous ftp or gopher: cpt.univ-mrs.fr

\end{titlepage}

The study of QCD current correlators at finite temperatures has been
of increased interest recently
(for an extensive review, see Ref.\cite{sh}).
By studying their $T$-dependence
one may get an insight on
the changes in the hadron spectrum as the temperature
approaches the point $T_c\approx 150$\, MeV
at which the hadron phase is expected to undergo chiral and
deconfining phase transitions.
One may try to extend the method of
operator product expansion (OPE)
from $T=0$ to the case of $T\neq 0$ and  use it to evaluate the
finite temperature correlators.
The $T$-dependence of non-perturbative condensates is due
to interactions with the heat bath.
Since at low $T$ the heat bath is dominated by pions,
it is possible to use soft pion methods
to obtain it\cite{leut,q4,eek,hkl}.
However to leading order $O(T^2)$ the temperature corrections
to correlators may be determined on the basis of current
algebra and PCAC without actually using OPE\cite{ls,e2,dei}.
In this order a mixing of correlators in parity and/or isospin takes
place\cite{dei,ei}.
The important point is that this mixing arises from direct scattering
of the heat bath pions on the currents and
no thermal mass shift is
generated in this order. (A general argument regarding the absence
of mass shifts in this order was given in Ref.\cite{ls}).
It has been shown\cite{q4,koi} that the
$T$-dependence of the relevant condensates
exactly matches the mixing of the correlators leaving
no room for mass shifts.

The condensates mentioned above are due to Lorentz scalar operators
in the OPE.
An important new feature of OPE at $T\neq 0$ is that operators
with non-zero Lorentz spin $s$
contribute to correlation functions because their average over the
heat bath is non-zero. They contribute to higher orders in $T^2$ and
are new non-perturbative parameters to be
determined\cite{sh,eek,hkl}. They cannot be obtained using
soft pion approach.

In a recent paper\cite{va} we have shown that order $T^4$
corrections to the vector and axial correlators may be
obtained using dispersion relations for the
amplitudes of deep inelastic scattering on pions and discussed their
relation to the lowest $s=2$ condensate, the energy-momentum tensor.
The results were interpreted in terms of $\rho$ and $a_1$
thermal mass shifts which turned out
to be negative. In the present paper we consider $T^4$ corrections for
correlators with quantum numbers of a nucleon and isobar. There is no
similarity with deep inelastic scattering in this case,
but we can use OPE with account of the $s=2$ operators to obtain
the corrections responsible for the mass shifts.

Let us consider the thermal correlation function defined as

\begin{equation}
C_{N}(q,T)=
i\int d^4 x e^{iqx}\langle T\{\eta (x),\bar{\eta} (0)\}\rangle _T
\label{c}
\end{equation}
where

\begin{equation}
\langle T\{\eta (x),\bar{\eta} (0)\}\rangle _T =
\frac{\sum_{n}\langle n|T\{\eta(x),\bar{\eta}(0)\}
e^{-H/T}|n\rangle}{\sum_{n}\langle n|e^{-H/T}|n\rangle},
\label{cn}
\end{equation}
Here $H$ is the QCD Hamiltonian, the sum is over all states of
the spectrum, and the nucleon (proton) current is taken to be\cite{bl}

\begin{equation}
\eta = (u^a C\gamma_{\mu}u^b )\gamma_5\gamma_{\mu}d^c\,\epsilon^{abc}
\label{nc}
\end{equation}
where $C=\gamma_2\gamma_0$ is the charge conjugation matrix.
It is assumed that $q^2$ is space-like, $Q^2=-q^2>0$, and $Q^2$
is much larger than a characteristic hadronic scale,
$Q^2\gg R_{c}^{-2}$,
where $R_c$ is the confinement radius, $R_{c}^{-1}\sim 0.5\:$GeV.

To determine the correlator $C_N$ at low $T$ one has to calculate
the matrix elements in Eq.(\ref{cn}) over pions.
In the chiral limit the pion momenta are of order $T$,
and the leading $O(T^2)$ corrections
are obtained neglecting the pion momentum in the one-pion matrix
elements. In this order using PCAC, current algebra and doing the
integral over the thermal pion phase space

\begin{equation}
\int\frac{d^{4}p}{(2\pi )^4}\,\frac{\pi\delta(p^2)}{e^{|{\bf p}|/T}-1}\,
\sum_{a}\,
\langle\pi^{a} (p) |T\{\eta(x),\bar{\eta}(0)\}|\pi^{a} (p)\rangle
\label{ps}
\end{equation}
gives the exact result\cite{e2,koi}

\begin{equation}
C_N (q,T)=(1-\xi) C_N (q,0) -\xi\gamma_5 C_N (q,0)\gamma_5
\label{mixn}
\end{equation}
where

\begin{equation}
\xi =T^2/16F^{2}_{\pi}
\label{xi}
\end{equation}
and $F_{\pi}=93$\, MeV is the pion decay constant.
The $T^2/F^{2}_{\pi}$ corrections are due to the
contact interactions of pions with the currents (Fig.1, $a$ and $b$).
There is no need to use OPE to calculate $C_N$ in this order.
Indeed, the $T$-dependences of $s=0$
condensates which contribute to this order
exactly satisfy Eq.(\ref{mixn})\cite{koi}.
Thus, in order $T^2$ the finite temperature correlator $C_N (q,T)$
is expressed in terms of the $T=0$ correlators.
This is similar to the mixing of the
vector and axial correlators\cite{dei}, but in the nucleon case
both $C_N$ and $\gamma_5 C_N \gamma_5$ are contributed by the
nucleon and its parity partner $N(1535)$.
The choice of the nucleon current in Eq.(\ref{nc})
allows to avoid the compensation of their contributions
to the chirality violating part of $C_N$\cite{ch}.
It is evident from Eq.(\ref{mixn}) that there can be no nucleon mass
shift in order $T^2$. The absence of the nucleon mass shift
in this order was also demonstrated in Ref.\cite{ls} by an
explicit calculation of the $\pi N$ self-energy correction to the
nucleon propagator at $T\neq 0$. The mass shift is expected to appear
in the next order, $T^4$.

The $T^4$ corrections to $C_N$ are due either to two-pion matrix
elements with zero pion momenta, or to one-pion matrix elements with
non-zero pion momentum\cite{va}. The former are of order
$T^4/F^{4}_{\pi}$.
They come from the contact interaction of pions with the currents
\newpage
\newsavebox{\uppercircle}
\savebox{\uppercircle}(0,25)[c]{
\put( 25.00,  0.00){\circle*{0.2}}
\put( 25.00,  0.50){\circle*{0.2}}
\put( 24.98,  1.00){\circle*{0.2}}
\put( 24.96,  1.50){\circle*{0.2}}
\put( 24.92,  2.00){\circle*{0.2}}
\put( 24.92,  2.00){\circle*{0.2}}
\put( 24.92,  2.00){\circle*{0.2}}
\put( 24.92,  2.00){\circle*{0.2}}
\put( 24.92,  2.00){\circle*{0.2}}
\put( 24.92,  2.00){\circle*{0.2}}
\put( 24.92,  2.00){\circle*{0.2}}
\put( 24.92,  2.00){\circle*{0.2}}
\put( 24.92,  2.00){\circle*{0.2}}
\put( 24.92,  2.00){\circle*{0.2}}
\put( 24.92,  2.00){\circle*{0.2}}
\put( 23.88,  7.39){\circle*{0.2}}
\put( 23.73,  7.86){\circle*{0.2}}
\put( 23.57,  8.34){\circle*{0.2}}
\put( 23.40,  8.81){\circle*{0.2}}
\put( 23.22,  9.27){\circle*{0.2}}
\put( 23.03,  9.74){\circle*{0.2}}
\put( 22.83, 10.19){\circle*{0.2}}
\put( 22.62, 10.65){\circle*{0.2}}
\put( 22.40, 11.10){\circle*{0.2}}
\put( 22.17, 11.54){\circle*{0.2}}
\put( 22.17, 11.54){\circle*{0.2}}
\put( 22.17, 11.54){\circle*{0.2}}
\put( 22.17, 11.54){\circle*{0.2}}
\put( 22.17, 11.54){\circle*{0.2}}
\put( 22.17, 11.54){\circle*{0.2}}
\put( 22.17, 11.54){\circle*{0.2}}
\put( 22.17, 11.54){\circle*{0.2}}
\put( 22.17, 11.54){\circle*{0.2}}
\put( 22.17, 11.54){\circle*{0.2}}
\put( 22.17, 11.54){\circle*{0.2}}
\put( 19.12, 16.11){\circle*{0.2}}
\put( 18.80, 16.48){\circle*{0.2}}
\put( 18.46, 16.86){\circle*{0.2}}
\put( 18.12, 17.22){\circle*{0.2}}
\put( 17.77, 17.58){\circle*{0.2}}
\put( 17.42, 17.93){\circle*{0.2}}
\put( 17.06, 18.28){\circle*{0.2}}
\put( 16.69, 18.62){\circle*{0.2}}
\put( 16.31, 18.95){\circle*{0.2}}
\put( 15.93, 19.27){\circle*{0.2}}
\put( 15.93, 19.27){\circle*{0.2}}
\put( 15.93, 19.27){\circle*{0.2}}
\put( 15.93, 19.27){\circle*{0.2}}
\put( 15.93, 19.27){\circle*{0.2}}
\put( 15.93, 19.27){\circle*{0.2}}
\put( 15.93, 19.27){\circle*{0.2}}
\put( 15.93, 19.27){\circle*{0.2}}
\put( 15.93, 19.27){\circle*{0.2}}
\put( 15.93, 19.27){\circle*{0.2}}
\put( 11.78, 22.05){\circle*{0.2}}
\put( 11.34, 22.28){\circle*{0.2}}
\put( 10.89, 22.50){\circle*{0.2}}
\put( 10.44, 22.72){\circle*{0.2}}
\put(  9.98, 22.92){\circle*{0.2}}
\put(  9.52, 23.12){\circle*{0.2}}
\put(  9.06, 23.30){\circle*{0.2}}
\put(  8.59, 23.48){\circle*{0.2}}
\put(  8.12, 23.64){\circle*{0.2}}
\put(  7.65, 23.80){\circle*{0.2}}
\put(  7.65, 23.80){\circle*{0.2}}
\put(  7.65, 23.80){\circle*{0.2}}
\put(  7.65, 23.80){\circle*{0.2}}
\put(  7.65, 23.80){\circle*{0.2}}
\put(  7.65, 23.80){\circle*{0.2}}
\put(  7.65, 23.80){\circle*{0.2}}
\put(  7.65, 23.80){\circle*{0.2}}
\put(  7.65, 23.80){\circle*{0.2}}
\put(  7.65, 23.80){\circle*{0.2}}
\put(  7.65, 23.80){\circle*{0.2}}
\put(  2.27, 24.90){\circle*{0.2}}
\put(  1.77, 24.94){\circle*{0.2}}
\put(  1.27, 24.97){\circle*{0.2}}
\put(  0.77, 24.99){\circle*{0.2}}
\put(  0.27, 25.00){\circle*{0.2}}
\put( -0.23, 25.00){\circle*{0.2}}
\put( -0.73, 24.99){\circle*{0.2}}
\put( -1.23, 24.97){\circle*{0.2}}
\put( -1.73, 24.94){\circle*{0.2}}
\put( -2.23, 24.90){\circle*{0.2}}
\put( -2.23, 24.90){\circle*{0.2}}
\put( -2.23, 24.90){\circle*{0.2}}
\put( -2.23, 24.90){\circle*{0.2}}
\put( -2.23, 24.90){\circle*{0.2}}
\put( -2.23, 24.90){\circle*{0.2}}
\put( -2.23, 24.90){\circle*{0.2}}
\put( -2.23, 24.90){\circle*{0.2}}
\put( -2.23, 24.90){\circle*{0.2}}
\put( -2.23, 24.90){\circle*{0.2}}
\put( -2.23, 24.90){\circle*{0.2}}
\put( -7.61, 23.81){\circle*{0.2}}
\put( -8.08, 23.66){\circle*{0.2}}
\put( -8.55, 23.49){\circle*{0.2}}
\put( -9.02, 23.32){\circle*{0.2}}
\put( -9.49, 23.13){\circle*{0.2}}
\put( -9.95, 22.94){\circle*{0.2}}
\put(-10.40, 22.73){\circle*{0.2}}
\put(-10.86, 22.52){\circle*{0.2}}
\put(-11.30, 22.30){\circle*{0.2}}
\put(-11.75, 22.07){\circle*{0.2}}
\put(-11.75, 22.07){\circle*{0.2}}
\put(-11.75, 22.07){\circle*{0.2}}
\put(-11.75, 22.07){\circle*{0.2}}
\put(-11.75, 22.07){\circle*{0.2}}
\put(-11.75, 22.07){\circle*{0.2}}
\put(-11.75, 22.07){\circle*{0.2}}
\put(-11.75, 22.07){\circle*{0.2}}
\put(-11.75, 22.07){\circle*{0.2}}
\put(-11.75, 22.07){\circle*{0.2}}
\put(-15.90, 19.29){\circle*{0.2}}
\put(-16.28, 18.97){\circle*{0.2}}
\put(-16.66, 18.64){\circle*{0.2}}
\put(-17.03, 18.31){\circle*{0.2}}
\put(-17.39, 17.96){\circle*{0.2}}
\put(-17.74, 17.61){\circle*{0.2}}
\put(-18.09, 17.25){\circle*{0.2}}
\put(-18.43, 16.89){\circle*{0.2}}
\put(-18.77, 16.51){\circle*{0.2}}
\put(-19.10, 16.14){\circle*{0.2}}
\put(-19.10, 16.14){\circle*{0.2}}
\put(-19.10, 16.14){\circle*{0.2}}
\put(-19.10, 16.14){\circle*{0.2}}
\put(-19.10, 16.14){\circle*{0.2}}
\put(-19.10, 16.14){\circle*{0.2}}
\put(-19.10, 16.14){\circle*{0.2}}
\put(-19.10, 16.14){\circle*{0.2}}
\put(-19.10, 16.14){\circle*{0.2}}
\put(-19.10, 16.14){\circle*{0.2}}
\put(-19.10, 16.14){\circle*{0.2}}
\put(-22.16, 11.58){\circle*{0.2}}
\put(-22.38, 11.13){\circle*{0.2}}
\put(-22.60, 10.68){\circle*{0.2}}
\put(-22.81, 10.23){\circle*{0.2}}
\put(-23.01,  9.77){\circle*{0.2}}
\put(-23.20,  9.31){\circle*{0.2}}
\put(-23.38,  8.84){\circle*{0.2}}
\put(-23.56,  8.37){\circle*{0.2}}
\put(-23.72,  7.90){\circle*{0.2}}
\put(-23.87,  7.43){\circle*{0.2}}
\put(-23.87,  7.43){\circle*{0.2}}
\put(-23.87,  7.43){\circle*{0.2}}
\put(-23.87,  7.43){\circle*{0.2}}
\put(-23.87,  7.43){\circle*{0.2}}
\put(-23.87,  7.43){\circle*{0.2}}
\put(-23.87,  7.43){\circle*{0.2}}
\put(-23.87,  7.43){\circle*{0.2}}
\put(-23.87,  7.43){\circle*{0.2}}
\put(-23.87,  7.43){\circle*{0.2}}
\put(-23.87,  7.43){\circle*{0.2}}
\put(-24.92,  2.04){\circle*{0.2}}
\put(-24.95,  1.54){\circle*{0.2}}
\put(-24.98,  1.04){\circle*{0.2}}
\put(-24.99,  0.54){\circle*{0.2}}
}

\newsavebox{\lowercircle}
\savebox{\lowercircle}(0,-25)[c]{
\put(-25.00,  0.04){\circle*{0.2}}
\put(-25.00, -0.46){\circle*{0.2}}
\put(-24.98, -0.96){\circle*{0.2}}
\put(-24.96, -1.46){\circle*{0.2}}
\put(-24.92, -1.96){\circle*{0.2}}
\put(-24.92, -1.96){\circle*{0.2}}
\put(-24.92, -1.96){\circle*{0.2}}
\put(-24.92, -1.96){\circle*{0.2}}
\put(-24.92, -1.96){\circle*{0.2}}
\put(-24.92, -1.96){\circle*{0.2}}
\put(-24.92, -1.96){\circle*{0.2}}
\put(-24.92, -1.96){\circle*{0.2}}
\put(-24.92, -1.96){\circle*{0.2}}
\put(-24.92, -1.96){\circle*{0.2}}
\put(-24.92, -1.96){\circle*{0.2}}
\put(-23.90, -7.35){\circle*{0.2}}
\put(-23.74, -7.83){\circle*{0.2}}
\put(-23.58, -8.30){\circle*{0.2}}
\put(-23.41, -8.77){\circle*{0.2}}
\put(-23.23, -9.24){\circle*{0.2}}
\put(-23.04, -9.70){\circle*{0.2}}
\put(-22.84,-10.16){\circle*{0.2}}
\put(-22.64,-10.61){\circle*{0.2}}
\put(-22.42,-11.06){\circle*{0.2}}
\put(-22.19,-11.51){\circle*{0.2}}
\put(-22.19,-11.51){\circle*{0.2}}
\put(-22.19,-11.51){\circle*{0.2}}
\put(-22.19,-11.51){\circle*{0.2}}
\put(-22.19,-11.51){\circle*{0.2}}
\put(-22.19,-11.51){\circle*{0.2}}
\put(-22.19,-11.51){\circle*{0.2}}
\put(-22.19,-11.51){\circle*{0.2}}
\put(-22.19,-11.51){\circle*{0.2}}
\put(-22.19,-11.51){\circle*{0.2}}
\put(-22.19,-11.51){\circle*{0.2}}
\put(-19.15,-16.07){\circle*{0.2}}
\put(-18.82,-16.45){\circle*{0.2}}
\put(-18.49,-16.83){\circle*{0.2}}
\put(-18.15,-17.19){\circle*{0.2}}
\put(-17.80,-17.55){\circle*{0.2}}
\put(-17.45,-17.91){\circle*{0.2}}
\put(-17.08,-18.25){\circle*{0.2}}
\put(-16.72,-18.59){\circle*{0.2}}
\put(-16.34,-18.92){\circle*{0.2}}
\put(-15.96,-19.24){\circle*{0.2}}
\put(-15.96,-19.24){\circle*{0.2}}
\put(-15.96,-19.24){\circle*{0.2}}
\put(-15.96,-19.24){\circle*{0.2}}
\put(-15.96,-19.24){\circle*{0.2}}
\put(-15.96,-19.24){\circle*{0.2}}
\put(-15.96,-19.24){\circle*{0.2}}
\put(-15.96,-19.24){\circle*{0.2}}
\put(-15.96,-19.24){\circle*{0.2}}
\put(-15.96,-19.24){\circle*{0.2}}
\put(-15.96,-19.24){\circle*{0.2}}
\put(-11.38,-22.26){\circle*{0.2}}
\put(-10.93,-22.49){\circle*{0.2}}
\put(-10.48,-22.70){\circle*{0.2}}
\put(-10.02,-22.90){\circle*{0.2}}
\put( -9.56,-23.10){\circle*{0.2}}
\put( -9.10,-23.29){\circle*{0.2}}
\put( -8.63,-23.46){\circle*{0.2}}
\put( -8.16,-23.63){\circle*{0.2}}
\put( -7.68,-23.79){\circle*{0.2}}
\put( -7.68,-23.79){\circle*{0.2}}
\put( -7.68,-23.79){\circle*{0.2}}
\put( -7.68,-23.79){\circle*{0.2}}
\put( -7.68,-23.79){\circle*{0.2}}
\put( -7.68,-23.79){\circle*{0.2}}
\put( -7.68,-23.79){\circle*{0.2}}
\put( -7.68,-23.79){\circle*{0.2}}
\put( -7.68,-23.79){\circle*{0.2}}
\put( -7.68,-23.79){\circle*{0.2}}
\put( -7.68,-23.79){\circle*{0.2}}
\put( -2.31,-24.89){\circle*{0.2}}
\put( -1.81,-24.93){\circle*{0.2}}
\put( -1.31,-24.97){\circle*{0.2}}
\put( -0.81,-24.99){\circle*{0.2}}
\put( -0.31,-25.00){\circle*{0.2}}
\put(  0.19,-25.00){\circle*{0.2}}
\put(  0.69,-24.99){\circle*{0.2}}
\put(  1.19,-24.97){\circle*{0.2}}
\put(  1.69,-24.94){\circle*{0.2}}
\put(  2.19,-24.90){\circle*{0.2}}
\put(  2.19,-24.90){\circle*{0.2}}
\put(  2.19,-24.90){\circle*{0.2}}
\put(  2.19,-24.90){\circle*{0.2}}
\put(  2.19,-24.90){\circle*{0.2}}
\put(  2.19,-24.90){\circle*{0.2}}
\put(  2.19,-24.90){\circle*{0.2}}
\put(  2.19,-24.90){\circle*{0.2}}
\put(  2.19,-24.90){\circle*{0.2}}
\put(  2.19,-24.90){\circle*{0.2}}
\put(  2.19,-24.90){\circle*{0.2}}
\put(  7.57,-23.83){\circle*{0.2}}
\put(  8.04,-23.67){\circle*{0.2}}
\put(  8.52,-23.50){\circle*{0.2}}
\put(  8.98,-23.33){\circle*{0.2}}
\put(  9.45,-23.15){\circle*{0.2}}
\put(  9.91,-22.95){\circle*{0.2}}
\put( 10.37,-22.75){\circle*{0.2}}
\put( 10.82,-22.54){\circle*{0.2}}
\put( 11.27,-22.32){\circle*{0.2}}
\put( 11.71,-22.09){\circle*{0.2}}
\put( 11.71,-22.09){\circle*{0.2}}
\put( 11.71,-22.09){\circle*{0.2}}
\put( 11.71,-22.09){\circle*{0.2}}
\put( 11.71,-22.09){\circle*{0.2}}
\put( 11.71,-22.09){\circle*{0.2}}
\put( 11.71,-22.09){\circle*{0.2}}
\put( 11.71,-22.09){\circle*{0.2}}
\put( 11.71,-22.09){\circle*{0.2}}
\put( 11.71,-22.09){\circle*{0.2}}
\put( 15.87,-19.32){\circle*{0.2}}
\put( 16.25,-19.00){\circle*{0.2}}
\put( 16.63,-18.67){\circle*{0.2}}
\put( 17.00,-18.33){\circle*{0.2}}
\put( 17.36,-17.99){\circle*{0.2}}
\put( 17.72,-17.64){\circle*{0.2}}
\put( 18.07,-17.28){\circle*{0.2}}
\put( 18.41,-16.92){\circle*{0.2}}
\put( 18.74,-16.54){\circle*{0.2}}
\put( 19.07,-16.17){\circle*{0.2}}
\put( 19.07,-16.17){\circle*{0.2}}
\put( 19.07,-16.17){\circle*{0.2}}
\put( 19.07,-16.17){\circle*{0.2}}
\put( 19.07,-16.17){\circle*{0.2}}
\put( 19.07,-16.17){\circle*{0.2}}
\put( 19.07,-16.17){\circle*{0.2}}
\put( 19.07,-16.17){\circle*{0.2}}
\put( 19.07,-16.17){\circle*{0.2}}
\put( 19.07,-16.17){\circle*{0.2}}
\put( 19.07,-16.17){\circle*{0.2}}
\put( 22.14,-11.62){\circle*{0.2}}
\put( 22.37,-11.17){\circle*{0.2}}
\put( 22.58,-10.72){\circle*{0.2}}
\put( 22.79,-10.27){\circle*{0.2}}
\put( 23.00, -9.81){\circle*{0.2}}
\put( 23.19, -9.35){\circle*{0.2}}
\put( 23.37, -8.88){\circle*{0.2}}
\put( 23.54, -8.41){\circle*{0.2}}
\put( 23.71, -7.94){\circle*{0.2}}
\put( 23.86, -7.46){\circle*{0.2}}
\put( 23.86, -7.46){\circle*{0.2}}
\put( 23.86, -7.46){\circle*{0.2}}
\put( 23.86, -7.46){\circle*{0.2}}
\put( 23.86, -7.46){\circle*{0.2}}
\put( 23.86, -7.46){\circle*{0.2}}
\put( 23.86, -7.46){\circle*{0.2}}
\put( 23.86, -7.46){\circle*{0.2}}
\put( 23.86, -7.46){\circle*{0.2}}
\put( 23.86, -7.46){\circle*{0.2}}
\put( 23.86, -7.46){\circle*{0.2}}
\put( 24.91, -2.08){\circle*{0.2}}
\put( 24.95, -1.58){\circle*{0.2}}
\put( 24.98, -1.08){\circle*{0.2}}
\put( 24.99, -0.58){\circle*{0.2}}
\put( 25.00, -0.08){\circle*{0.2}}
}

\noindent
\begin{picture}(425,560)(-115,-40)

\thicklines

\put( -50,502){\line(1,0){50}}
\put( -50,498){\line(1,0){50}}
\put( -50,498){\line(0,1){4}}
\put(  25,500){\usebox{\uppercircle}}
\put(  25,500){\usebox{\lowercircle}}
\put(  25,520){\line(0,1){10}}
\put(   0,450){\makebox(0,0)[c]{(a)}}

\put( 175,502){\line(1,0){50}}
\put( 175,498){\line(1,0){50}}
\put( 175,498){\line(0,1){2}}
\put( 225,498){\line(0,1){2}}
\put( 200,520){\line(0,1){10}}
\put( 200,500){\usebox{\uppercircle}}
\put( 200,450){\makebox(0,0)[c]{(b)}}

\put(-50,350){\usebox{\uppercircle}}
\put(-50,350){\usebox{\lowercircle}}
\put(-50,370){\line(0,1){10}}
\put(-25,352){\line(1,0){50}}
\put(-25,348){\line(1,0){50}}
\put( 50,350){\usebox{\uppercircle}}
\put( 50,350){\usebox{\lowercircle}}
\put( 50,370){\line(0,1){10}}
\put(  0,275){\makebox(0,0)[c]{(c)}}

\put( 175,352){\line(1,0){50}}
\put( 175,348){\line(1,0){50}}
\put( 175,348){\line(0,1){4}}
\put( 225,348){\line(0,1){4}}
\put( 225,397){\line(0,1){10}}
\put( 225,293){\line(0,1){10}}
\put( 225,377){\usebox{\uppercircle}}
\put( 225,377){\usebox{\lowercircle}}
\put( 225,323){\usebox{\uppercircle}}
\put( 225,323){\usebox{\lowercircle}}
\put( 200,275){\makebox(0,0)[c]{(d)}}

\put( -25,202){\line(1,0){50}}
\put( -25,198){\line(1,0){50}}
\put(   0,200){\usebox{\uppercircle}}
\put(   0,200){\usebox{\lowercircle}}
\put(   0,220){\line(0,1){10}}
\put(   0,170){\line(0,1){10}}
\put(   0,150){\makebox(0,0)[c]{(e)}}

\put( 150,202){\line(1,0){50}}
\put( 150,198){\line(1,0){50}}
\put( 150,198){\line(0,1){4}}
\put( 175,200){\usebox{\uppercircle}}
\put( 175,220){\line(0,1){10}}
\put( 225,200){\usebox{\uppercircle}}
\put( 225,200){\usebox{\lowercircle}}
\put( 225,220){\line(0,1){10}}
\put( 200,150){\makebox(0,0)[c]{(f)}}

\put( -25,  0){\line(1,0){50}}
\put( -25, -4){\line(1,0){50}}
\put( -25, -4){\line(0,1){4}}
\put(  25, -4){\line(0,1){4}}
\put( -25, 25){\usebox{\uppercircle}}
\put( -25, 25){\usebox{\lowercircle}}
\put(  -5, 25){\line(1,0){10}}
\put( -25, 75){\usebox{\uppercircle}}
\put( -25, 75){\usebox{\lowercircle}}
\put( -25, 50){\circle*{5}}
\put( -25, 95){\line(0,1){10}}
\put(   0,-25){\makebox(0,0)[c]{(g)}}

\put( 175,  0){\line(1,0){50}}
\put( 175, -4){\line(1,0){50}}
\put( 175, -4){\line(0,1){4}}
\put( 225, -4){\line(0,1){4}}
\put( 200,  0){\usebox{\uppercircle}}
\multiput( 214, 14)(0.25,0.25){30}{\circle*{0.2}}
\put( 200, 50){\usebox{\uppercircle}}
\put( 200, 50){\usebox{\lowercircle}}
\put( 200, 25){\circle*{5}}
\put( 200, 70){\line(0,1){10}}
\put( 200,-25){\makebox(0,0)[c]{(h)}}


\end{picture}

\noindent
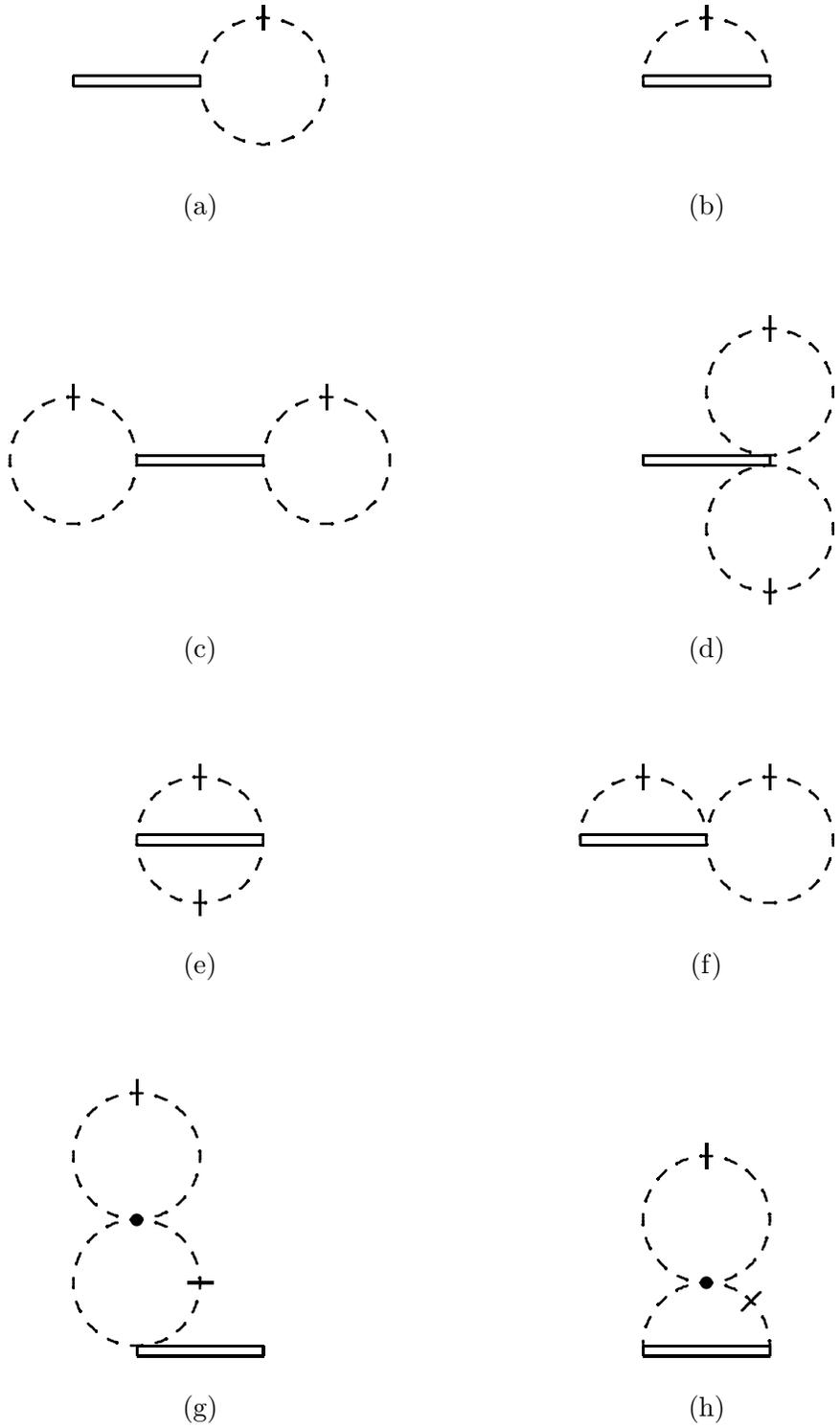
Fig. 1. Contact terms to two loops. Double lines are $T=0$ correlators,
crossed dashed lines denote thermal pions.
\newpage
\noindent
in two loops (Fig.1, $c$ through $h$), can be calculated without OPE,
and contribute to the change in
the coupling of the nucleon to the current (see Eq.(\ref{gamma}) and
(\ref{1}) below).
The latter are of order $T^4/Q^4$, they arise from
the pion scattering on the intermediate state in the
correlator and may be interpreted in terms of a thermal mass shift.
They will also break the Lorentz covariance
preserved in order $T^2$ as is seen from Eq.(\ref{mixn}) in which
$C_N$ has the form

\begin{equation}
C_N (q^2,T)=C_1 (q^2,T)\hat{q} +C_2 (q^2,T)
\end{equation}
where $\hat{q}=q_{\mu}\gamma_{\mu}$.
The general form of the
correlator in the rest frame of the heat bath is given by

\begin{equation}
C_N (q,T)=C_1 (q,T)\gamma_0 q_0 +
C^{\prime}_2 (q,T)\mbox{\boldmath $\gamma q $} +C_2 (q,T)
\end{equation}
where $C^{\prime}_1\neq -C_1$ and all $C$'s depend separately on $q_0$
and $|{\bf q}|$. We will consider the case of ${\bf q} =0$, $q_0\neq 0$,
when we are left with the
same number of independent structures as at $T=0$.

In terms of OPE, there are two $s=2$ operators in the leading twist
which provide the $T^4/Q^4$ corrections to the correlator.
They are the quark and gluon energy-momentum tensors

\begin{eqnarray}
\theta_{\mu\nu}^q &=&\frac{i}{2}
(\bar{q}\gamma_{\mu}D_{\nu}q+\bar{q}\gamma_{\nu}D_{\mu}q),
{}~~~(q=u\, ,\, d)
\nonumber \\
\theta_{\mu\nu}^G &=&G_{\mu\alpha}^{a}G_{\alpha\nu}^{a}-
\frac{1}{4}g_{\mu\nu}G_{\alpha\beta}^{a}G_{\beta\alpha}^{a}\,
\label{theta}
\end{eqnarray}
normalized according to

\begin{equation}
\langle \pi (p)|\theta_{\mu\nu}^u +\theta_{\mu\nu}^d
+\theta_{\mu\nu}^G|\pi (p)\rangle =2p_{\mu}p_{\nu}
\label{tot}
\end{equation}
where $\langle \pi (p)|\pi (p^{\prime})\rangle = (2\pi)^3\, 2E\,
\delta^{(3)}({\bf p}-{\bf p}^{\prime})$.
The contributions of the quark and gluon tensors to Eq.(\ref{tot})
are about the same\cite{bb,eek}.
However, the gluon tensor enters OPE for the correlator being
multiplied by $\alpha_{s}(Q^2)$ and we will neglect its contribution.
In what follows we will also omit the anomalous dimension of
$\theta_{\mu\nu}^{q}$.

The calculation of the quark tensor contribution to the correlator
can most easily be done in the coordinate representation by cutting
quark lines and using the second term in the expansion of the quark
operator, $q(x)=q(0)+x_{\lambda}D_{\lambda}q(0)+...$.


Cutting one quark line gives the chirality conserving part.
In case of $d$-quark this results in matrix elements
of the form $\langle\pi (p)|\bar{d}D_{\lambda}d|\pi (p)\rangle$ which
are proportional to the

\noindent
\begin{picture}(425,140)(-62,20)


\thicklines

\put(0,100){\line(1,0){100}}
\put(0,100){\line(1,1){30}}
\put(100,100){\line(-1,1){30}}
\put(80,115){\line(0,1){10}}
\put(75,120){\line(1,0){10}}
\put(-10,100){\makebox(0,0)[c]{$0$}}
\put(110,100){\makebox(0,0)[c]{$x$}}
\put(100,120){\makebox(0,0)[c]{$u$}}
\put(50,110){\makebox(0,0)[c]{$u$}}
\put(50,65){\makebox(0,0)[c]{$d$}}

\put(50,30){\makebox(0,0)[c]{(a)}}

\put(200,100){\line(1,0){100}}
\put(200,100){\line(1,1){30}}
\put(300,100){\line(-1,1){30}}
\put(200,100){\line(1,-1){30}}
\put(300,100){\line(-1,-1){30}}
\put(280,115){\line(0,1){10}}
\put(275,120){\line(1,0){10}}
\put(190,100){\makebox(0,0)[c]{$0$}}
\put(310,100){\makebox(0,0)[c]{$x$}}
\put(300,120){\makebox(0,0)[c]{$u$}}
\put(250,110){\makebox(0,0)[c]{$u$}}
\put(290,75){\makebox(0,0)[c]{$d$}}

\put(250,30){\makebox(0,0)[c]{(b)}}

\put( -0.09,100.11){\circle*{0.2}}
\put(  0.16, 99.86){\circle*{0.2}}
\put(  0.41, 99.61){\circle*{0.2}}
\put(  0.66, 99.36){\circle*{0.2}}
\put(  0.92, 99.12){\circle*{0.2}}
\put(  1.17, 98.87){\circle*{0.2}}
\put(  1.43, 98.63){\circle*{0.2}}
\put(  1.68, 98.38){\circle*{0.2}}
\put(  1.94, 98.14){\circle*{0.2}}
\put(  2.20, 97.90){\circle*{0.2}}
\put(  2.46, 97.67){\circle*{0.2}}
\put(  2.73, 97.43){\circle*{0.2}}
\put(  2.99, 97.19){\circle*{0.2}}
\put(  3.25, 96.96){\circle*{0.2}}
\put(  3.52, 96.73){\circle*{0.2}}
\put(  3.79, 96.49){\circle*{0.2}}
\put(  4.06, 96.26){\circle*{0.2}}
\put(  4.32, 96.03){\circle*{0.2}}
\put(  4.60, 95.81){\circle*{0.2}}
\put(  4.87, 95.58){\circle*{0.2}}
\put(  5.14, 95.36){\circle*{0.2}}
\put(  5.41, 95.13){\circle*{0.2}}
\put(  5.69, 94.91){\circle*{0.2}}
\put(  5.96, 94.69){\circle*{0.2}}
\put(  6.24, 94.47){\circle*{0.2}}
\put(  6.52, 94.25){\circle*{0.2}}
\put(  6.80, 94.03){\circle*{0.2}}
\put(  7.08, 93.82){\circle*{0.2}}
\put(  7.36, 93.61){\circle*{0.2}}
\put(  7.64, 93.39){\circle*{0.2}}
\put(  7.93, 93.18){\circle*{0.2}}
\put(  8.21, 92.97){\circle*{0.2}}
\put(  8.50, 92.76){\circle*{0.2}}
\put(  8.78, 92.56){\circle*{0.2}}
\put(  9.07, 92.35){\circle*{0.2}}
\put(  9.36, 92.15){\circle*{0.2}}
\put(  9.65, 91.95){\circle*{0.2}}
\put(  9.94, 91.74){\circle*{0.2}}
\put( 10.23, 91.54){\circle*{0.2}}
\put( 10.53, 91.35){\circle*{0.2}}
\put( 10.82, 91.15){\circle*{0.2}}
\put( 11.11, 90.95){\circle*{0.2}}
\put( 11.41, 90.76){\circle*{0.2}}
\put( 11.71, 90.57){\circle*{0.2}}
\put( 12.00, 90.38){\circle*{0.2}}
\put( 12.30, 90.19){\circle*{0.2}}
\put( 12.60, 90.00){\circle*{0.2}}
\put( 12.90, 89.81){\circle*{0.2}}
\put( 13.20, 89.63){\circle*{0.2}}
\put( 13.51, 89.45){\circle*{0.2}}
\put( 13.81, 89.27){\circle*{0.2}}
\put( 14.11, 89.09){\circle*{0.2}}
\put( 14.42, 88.91){\circle*{0.2}}
\put( 14.72, 88.73){\circle*{0.2}}
\put( 15.03, 88.55){\circle*{0.2}}
\put( 15.34, 88.38){\circle*{0.2}}
\put( 15.65, 88.21){\circle*{0.2}}
\put( 15.96, 88.04){\circle*{0.2}}
\put( 16.27, 87.87){\circle*{0.2}}
\put( 16.58, 87.70){\circle*{0.2}}
\put( 16.89, 87.53){\circle*{0.2}}
\put( 17.20, 87.37){\circle*{0.2}}
\put( 17.52, 87.20){\circle*{0.2}}
\put( 17.83, 87.04){\circle*{0.2}}
\put( 18.15, 86.88){\circle*{0.2}}
\put( 18.46, 86.72){\circle*{0.2}}
\put( 18.78, 86.57){\circle*{0.2}}
\put( 19.10, 86.41){\circle*{0.2}}
\put( 19.41, 86.26){\circle*{0.2}}
\put( 19.73, 86.11){\circle*{0.2}}
\put( 20.05, 85.96){\circle*{0.2}}
\put( 20.37, 85.81){\circle*{0.2}}
\put( 20.70, 85.66){\circle*{0.2}}
\put( 21.02, 85.51){\circle*{0.2}}
\put( 21.34, 85.37){\circle*{0.2}}
\put( 21.66, 85.23){\circle*{0.2}}
\put( 21.99, 85.09){\circle*{0.2}}
\put( 22.31, 84.95){\circle*{0.2}}
\put( 22.64, 84.81){\circle*{0.2}}
\put( 22.96, 84.67){\circle*{0.2}}
\put( 23.29, 84.54){\circle*{0.2}}
\put( 23.62, 84.41){\circle*{0.2}}
\put( 23.95, 84.28){\circle*{0.2}}
\put( 24.28, 84.15){\circle*{0.2}}
\put( 24.61, 84.02){\circle*{0.2}}
\put( 24.94, 83.89){\circle*{0.2}}
\put( 25.27, 83.77){\circle*{0.2}}
\put( 25.60, 83.64){\circle*{0.2}}
\put( 25.93, 83.52){\circle*{0.2}}
\put( 26.26, 83.40){\circle*{0.2}}
\put( 26.60, 83.29){\circle*{0.2}}
\put( 26.93, 83.17){\circle*{0.2}}
\put( 27.26, 83.06){\circle*{0.2}}
\put( 27.60, 82.94){\circle*{0.2}}
\put( 27.94, 82.83){\circle*{0.2}}
\put( 28.27, 82.72){\circle*{0.2}}
\put( 28.61, 82.61){\circle*{0.2}}
\put( 28.95, 82.51){\circle*{0.2}}
\put( 29.28, 82.40){\circle*{0.2}}
\put( 29.62, 82.30){\circle*{0.2}}
\put( 29.96, 82.20){\circle*{0.2}}
\put( 30.30, 82.10){\circle*{0.2}}
\put( 30.64, 82.00){\circle*{0.2}}
\put( 30.98, 81.91){\circle*{0.2}}
\put( 31.32, 81.81){\circle*{0.2}}
\put( 31.66, 81.72){\circle*{0.2}}
\put( 32.00, 81.63){\circle*{0.2}}
\put( 32.34, 81.54){\circle*{0.2}}
\put( 32.69, 81.45){\circle*{0.2}}
\put( 33.03, 81.37){\circle*{0.2}}
\put( 33.37, 81.28){\circle*{0.2}}
\put( 33.72, 81.20){\circle*{0.2}}
\put( 34.06, 81.12){\circle*{0.2}}
\put( 34.41, 81.04){\circle*{0.2}}
\put( 34.75, 80.96){\circle*{0.2}}
\put( 35.10, 80.89){\circle*{0.2}}
\put( 35.44, 80.81){\circle*{0.2}}
\put( 35.79, 80.74){\circle*{0.2}}
\put( 36.14, 80.67){\circle*{0.2}}
\put( 36.48, 80.60){\circle*{0.2}}
\put( 36.83, 80.54){\circle*{0.2}}
\put( 37.18, 80.47){\circle*{0.2}}
\put( 37.52, 80.41){\circle*{0.2}}
\put( 37.87, 80.35){\circle*{0.2}}
\put( 38.22, 80.29){\circle*{0.2}}
\put( 38.57, 80.23){\circle*{0.2}}
\put( 38.92, 80.17){\circle*{0.2}}
\put( 39.27, 80.12){\circle*{0.2}}
\put( 39.62, 80.07){\circle*{0.2}}
\put( 39.97, 80.02){\circle*{0.2}}
\put( 40.32, 79.97){\circle*{0.2}}
\put( 40.67, 79.92){\circle*{0.2}}
\put( 41.02, 79.87){\circle*{0.2}}
\put( 41.37, 79.83){\circle*{0.2}}
\put( 41.72, 79.79){\circle*{0.2}}
\put( 42.07, 79.75){\circle*{0.2}}
\put( 42.42, 79.71){\circle*{0.2}}
\put( 42.77, 79.67){\circle*{0.2}}
\put( 43.13, 79.64){\circle*{0.2}}
\put( 43.48, 79.60){\circle*{0.2}}
\put( 43.83, 79.57){\circle*{0.2}}
\put( 44.18, 79.54){\circle*{0.2}}
\put( 44.53, 79.51){\circle*{0.2}}
\put( 44.89, 79.49){\circle*{0.2}}
\put( 45.24, 79.46){\circle*{0.2}}
\put( 45.59, 79.44){\circle*{0.2}}
\put( 45.94, 79.42){\circle*{0.2}}
\put( 46.30, 79.40){\circle*{0.2}}
\put( 46.65, 79.38){\circle*{0.2}}
\put( 47.00, 79.36){\circle*{0.2}}
\put( 47.36, 79.35){\circle*{0.2}}
\put( 47.71, 79.34){\circle*{0.2}}
\put( 48.06, 79.33){\circle*{0.2}}
\put( 48.42, 79.32){\circle*{0.2}}
\put( 48.77, 79.31){\circle*{0.2}}
\put( 49.12, 79.31){\circle*{0.2}}
\put( 49.48, 79.30){\circle*{0.2}}
\put( 49.83, 79.30){\circle*{0.2}}
\put( 50.18, 79.30){\circle*{0.2}}
\put( 50.54, 79.30){\circle*{0.2}}
\put( 50.89, 79.31){\circle*{0.2}}
\put( 51.25, 79.31){\circle*{0.2}}
\put( 51.60, 79.32){\circle*{0.2}}
\put( 51.95, 79.33){\circle*{0.2}}
\put( 52.31, 79.34){\circle*{0.2}}
\put( 52.66, 79.35){\circle*{0.2}}
\put( 53.01, 79.36){\circle*{0.2}}
\put( 53.36, 79.38){\circle*{0.2}}
\put( 53.72, 79.40){\circle*{0.2}}
\put( 54.07, 79.42){\circle*{0.2}}
\put( 54.42, 79.44){\circle*{0.2}}
\put( 54.78, 79.46){\circle*{0.2}}
\put( 55.13, 79.49){\circle*{0.2}}
\put( 55.48, 79.51){\circle*{0.2}}
\put( 55.83, 79.54){\circle*{0.2}}
\put( 56.19, 79.57){\circle*{0.2}}
\put( 56.54, 79.60){\circle*{0.2}}
\put( 56.89, 79.64){\circle*{0.2}}
\put( 57.24, 79.67){\circle*{0.2}}
\put( 57.59, 79.71){\circle*{0.2}}
\put( 57.94, 79.75){\circle*{0.2}}
\put( 58.30, 79.79){\circle*{0.2}}
\put( 58.65, 79.83){\circle*{0.2}}
\put( 59.00, 79.87){\circle*{0.2}}
\put( 59.35, 79.92){\circle*{0.2}}
\put( 59.70, 79.97){\circle*{0.2}}
\put( 60.05, 80.02){\circle*{0.2}}
\put( 60.40, 80.07){\circle*{0.2}}
\put( 60.75, 80.12){\circle*{0.2}}
\put( 61.10, 80.18){\circle*{0.2}}
\put( 61.45, 80.23){\circle*{0.2}}
\put( 61.79, 80.29){\circle*{0.2}}
\put( 62.14, 80.35){\circle*{0.2}}
\put( 62.49, 80.41){\circle*{0.2}}
\put( 62.84, 80.48){\circle*{0.2}}
\put( 63.19, 80.54){\circle*{0.2}}
\put( 63.53, 80.61){\circle*{0.2}}
\put( 63.88, 80.68){\circle*{0.2}}
\put( 64.23, 80.75){\circle*{0.2}}
\put( 64.57, 80.82){\circle*{0.2}}
\put( 64.92, 80.89){\circle*{0.2}}
\put( 65.26, 80.97){\circle*{0.2}}
\put( 65.61, 81.04){\circle*{0.2}}
\put( 65.95, 81.12){\circle*{0.2}}
\put( 66.30, 81.20){\circle*{0.2}}
\put( 66.64, 81.29){\circle*{0.2}}
\put( 66.98, 81.37){\circle*{0.2}}
\put( 67.33, 81.46){\circle*{0.2}}
\put( 67.67, 81.54){\circle*{0.2}}
\put( 68.01, 81.63){\circle*{0.2}}
\put( 68.35, 81.72){\circle*{0.2}}
\put( 68.70, 81.82){\circle*{0.2}}
\put( 69.04, 81.91){\circle*{0.2}}
\put( 69.38, 82.01){\circle*{0.2}}
\put( 69.72, 82.10){\circle*{0.2}}
\put( 70.05, 82.20){\circle*{0.2}}
\put( 70.39, 82.31){\circle*{0.2}}
\put( 70.73, 82.41){\circle*{0.2}}
\put( 71.07, 82.51){\circle*{0.2}}
\put( 71.41, 82.62){\circle*{0.2}}
\put( 71.74, 82.73){\circle*{0.2}}
\put( 72.08, 82.84){\circle*{0.2}}
\put( 72.42, 82.95){\circle*{0.2}}
\put( 72.75, 83.06){\circle*{0.2}}
\put( 73.08, 83.17){\circle*{0.2}}
\put( 73.42, 83.29){\circle*{0.2}}
\put( 73.75, 83.41){\circle*{0.2}}
\put( 74.08, 83.53){\circle*{0.2}}
\put( 74.42, 83.65){\circle*{0.2}}
\put( 74.75, 83.77){\circle*{0.2}}
\put( 75.08, 83.90){\circle*{0.2}}
\put( 75.41, 84.02){\circle*{0.2}}
\put( 75.74, 84.15){\circle*{0.2}}
\put( 76.07, 84.28){\circle*{0.2}}
\put( 76.40, 84.41){\circle*{0.2}}
\put( 76.72, 84.54){\circle*{0.2}}
\put( 77.05, 84.68){\circle*{0.2}}
\put( 77.38, 84.82){\circle*{0.2}}
\put( 77.70, 84.95){\circle*{0.2}}
\put( 78.03, 85.09){\circle*{0.2}}
\put( 78.35, 85.23){\circle*{0.2}}
\put( 78.67, 85.38){\circle*{0.2}}
\put( 79.00, 85.52){\circle*{0.2}}
\put( 79.32, 85.67){\circle*{0.2}}
\put( 79.64, 85.81){\circle*{0.2}}
\put( 79.96, 85.96){\circle*{0.2}}
\put( 80.28, 86.11){\circle*{0.2}}
\put( 80.60, 86.27){\circle*{0.2}}
\put( 80.92, 86.42){\circle*{0.2}}
\put( 81.24, 86.57){\circle*{0.2}}
\put( 81.55, 86.73){\circle*{0.2}}
\put( 81.87, 86.89){\circle*{0.2}}
\put( 82.18, 87.05){\circle*{0.2}}
\put( 82.50, 87.21){\circle*{0.2}}
\put( 82.81, 87.37){\circle*{0.2}}
\put( 83.12, 87.54){\circle*{0.2}}
\put( 83.44, 87.71){\circle*{0.2}}
\put( 83.75, 87.87){\circle*{0.2}}
\put( 84.06, 88.04){\circle*{0.2}}
\put( 84.37, 88.21){\circle*{0.2}}
\put( 84.68, 88.39){\circle*{0.2}}
\put( 84.98, 88.56){\circle*{0.2}}
\put( 85.29, 88.74){\circle*{0.2}}
\put( 85.60, 88.91){\circle*{0.2}}
\put( 85.90, 89.09){\circle*{0.2}}
\put( 86.20, 89.27){\circle*{0.2}}
\put( 86.51, 89.46){\circle*{0.2}}
\put( 86.81, 89.64){\circle*{0.2}}
\put( 87.11, 89.82){\circle*{0.2}}
\put( 87.41, 90.01){\circle*{0.2}}
\put( 87.71, 90.20){\circle*{0.2}}
\put( 88.01, 90.39){\circle*{0.2}}
\put( 88.31, 90.58){\circle*{0.2}}
\put( 88.60, 90.77){\circle*{0.2}}
\put( 88.90, 90.96){\circle*{0.2}}
\put( 89.19, 91.16){\circle*{0.2}}
\put( 89.49, 91.36){\circle*{0.2}}
\put( 89.78, 91.55){\circle*{0.2}}
\put( 90.07, 91.75){\circle*{0.2}}
\put( 90.36, 91.95){\circle*{0.2}}
\put( 90.65, 92.16){\circle*{0.2}}
\put( 90.94, 92.36){\circle*{0.2}}
\put( 91.23, 92.57){\circle*{0.2}}
\put( 91.52, 92.77){\circle*{0.2}}
\put( 91.80, 92.98){\circle*{0.2}}
\put( 92.09, 93.19){\circle*{0.2}}
\put( 92.37, 93.40){\circle*{0.2}}
\put( 92.65, 93.61){\circle*{0.2}}
\put( 92.93, 93.83){\circle*{0.2}}
\put( 93.21, 94.04){\circle*{0.2}}
\put( 93.49, 94.26){\circle*{0.2}}
\put( 93.77, 94.48){\circle*{0.2}}
\put( 94.05, 94.70){\circle*{0.2}}
\put( 94.32, 94.92){\circle*{0.2}}
\put( 94.60, 95.14){\circle*{0.2}}
\put( 94.87, 95.37){\circle*{0.2}}
\put( 95.15, 95.59){\circle*{0.2}}
\put( 95.42, 95.82){\circle*{0.2}}
\put( 95.69, 96.04){\circle*{0.2}}
\put( 95.96, 96.27){\circle*{0.2}}
\put( 96.22, 96.50){\circle*{0.2}}
\put( 96.49, 96.74){\circle*{0.2}}
\put( 96.76, 96.97){\circle*{0.2}}
\put( 97.02, 97.20){\circle*{0.2}}
\put( 97.29, 97.44){\circle*{0.2}}
\put( 97.55, 97.68){\circle*{0.2}}
\put( 97.81, 97.91){\circle*{0.2}}
\put( 98.07, 98.15){\circle*{0.2}}
\put( 98.33, 98.40){\circle*{0.2}}
\put( 98.58, 98.64){\circle*{0.2}}
\put( 98.84, 98.88){\circle*{0.2}}
\put( 99.10, 99.13){\circle*{0.2}}
\put( 99.35, 99.37){\circle*{0.2}}
\put( 99.60, 99.62){\circle*{0.2}}
\put( 99.85, 99.87){\circle*{0.2}}

\end{picture}

\noindent
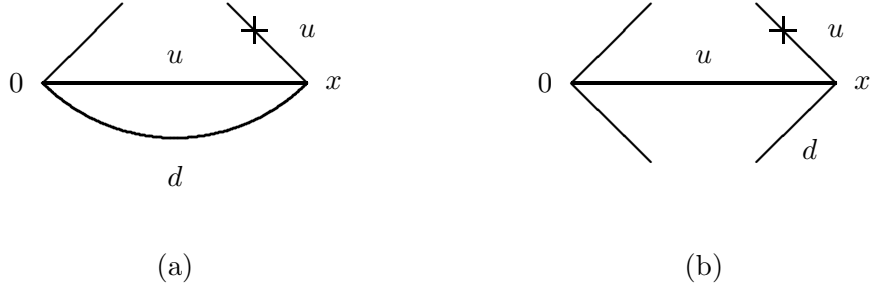
Fig. 2. Leading terms in OPE related to the nucleon thermal mass shift.
The quark leg with a cross corresponds to $x_{\lambda}D_{\lambda}q(0)$.

\vspace{1 cm}

\noindent
pion momentum $p$ and vanish under the
integration in Eq.(\ref{ps}). Cutting a $u$-quark line (Fig.2$a$) gives

\begin{equation}
i\frac{48M_2}{\pi ^{4}x^{8}}
(\hat{p} (px)x^{2}+\hat{x}(px)^{2})
\label{gammax}
\end{equation}
where $M_2$ is defined according to
$\langle \pi (p)|\theta_{\mu\nu}^u +\theta_{\mu\nu}^d |\pi (p)\rangle =
8M_2 p_{\mu}p_{\nu}$ and is equal to the second moment of the pion
structure function $M_2=\int_{0}^{1}F_{2}(x)dx$\cite{va}.

Integrating over the thermal pion phase space in the rest frame of the
heat bath with the help of

\begin{equation}
\int\frac{d^3 p}{(2\pi)^3\; 2|{\bf p}|}\;
\frac{p_{\mu}p_{\nu}}{{\rm exp}(|{\bf p}|)-1}=
(-g_{\mu\nu}+4g_{\mu 0}g_{\nu 0})\frac{\pi^2 T^4}{180}
\label{munu}
\end{equation}
and going over to the
momentum representation, we get the expression
to order $T^4$ for the chirality conserving structure
proportional to $\gamma_{0}q_0$

\begin{equation}
C_{1}(Q^2,T) = C_{1}(Q^2,0)
-\frac{4M_2}{15}\, T^4 \ln Q^2
\label{gamma}
\end{equation}
(order $\xi$ and  $\xi^2$ contact terms cancel in this structure).

Similarly, the chirality violating part of the $T^4/Q^4$ correction is
obtained by cutting two quark lines. Here, cutting the two $u$-quark
lines gives terms linear in the pion momentum which vanish under the
integration over pions, while cutting the $d$-quark and a $u$-quark
line (Fig.2$b$) gives terms quadratic in the pion momentum

\begin{equation}
-\frac{24M_2 (px)^{2}}{3\pi ^{2}x^{4}}\langle\bar{q}q\rangle
\label{1x}
\end{equation}
After integration over pions, this gives in the momentum representation

\begin{equation}
C_{2}(Q^2,T) = \left( 1-2\xi -\frac{2}{3}\xi^2\right) C_{2}(Q^2,0)
-\frac{2\pi ^{2}M_2\langle\bar{q}q\rangle}{5}\,\frac{T^4}{Q^2}
\label{1}
\end{equation}
Here $\langle\bar{q}q\rangle = -(240 MeV)^{3}$ is
the usual $T=0$ quark
condensate
\footnote{In deriving Eq.(\ref{1x}) we used the following
approximation for the matrix element over pion
$\langle\pi (p)|\bar{u}\gamma_{\mu}D_{\nu}u\,\bar{u}u|\pi (p)\rangle =
\langle\pi (p)|\bar{u}\gamma_{\mu}D_{\nu}u|\pi (p)\rangle
\langle\bar{u}u\rangle$. An estimate of the accuracy of this
factorization may be obtained by inserting a pion intermediate state.
Using the scale of OPE $\mu\sim 0.5$ Gev as the cut-off
in the integral over the momentum, we get that this contribution is
suppressed as $\mu ^{2}/8\pi ^{2}F_{\pi}^{2}\approx 0.3$.}.
In Eq.(\ref{gamma}) and (\ref{1}) we have taken into account
order $T^4/F^{4}_{\pi}$ corrections in a way similar to the case of
vector and axial correlators\cite{va}. The full answer was again
obtained by taking into account the initial (finite) state
interaction between two pions (Fig.2, $g$ and $h$).
The leading term in OPE for $C_2 (Q^2,0)$ is proportional to
$\langle\bar{q}q\rangle$.
A straightforward check shows that the $\xi^{2}$ term
in Eq.(\ref{1}) agrees with the
known\cite{leut} $T^4$ correction to the quark condensate.
It is easy to see that the last terms in Eqs.(\ref{gamma}) and
(\ref{1}) are indeed of order $T^4/Q^4$ compared to
$C_1 (Q^2, 0)$ and $C_2 (Q^2, 0)$, respectively.

Now, we would like to interpret
the $T^4$ corrections to the correlators in
Eqs.(\ref{gamma}) and (\ref{1})
in terms of the nucleon thermal mass shift.
The functions $C_1$ and $C_2$ may be presented in the form of
dispersion integrals

\begin{equation}
C_i (Q^2 ,T)=\frac{1}{\pi}\int_{0}^{\infty}
\frac{{\rm Im}C_i (s,T)ds}{s+Q^{2}}
\label{disp}
\end{equation}
where a sufficient number of subtractions is
implied and a standard model (lowest resonance + continuum)
for the spectral densities is usually assumed

\begin{equation}
{\rm Im}C_i (s,T)=\pi\lambda_{i}^{2}(T)\delta (s-m_{N}^{2}(T))+
\theta (s-s_0 (T))\rho (s,T)
\label{model}
\end{equation}
At $T=0$ the nucleon pole contribution to the correlator is

\begin{equation}
C_{N}^{(pole)} (Q^2,0)=
\lambda_{N} ^{2}\frac{\gamma_0 q_0 +m_N }{Q^2 +m_{N}^{2}}
\end{equation}
where $\lambda_N =\lambda_1 (0)=\lambda_2 (0)$ is the coupling of
the nucleon to the current,

\begin{equation}
\langle 0|\eta |N\rangle =\lambda_N v
\end{equation}
and the nucleon spinor $v$ is normalized as $\bar{v} v=2m_N$.
At $T\neq 0$ the general form of the nucleon propagator is

\begin{equation}
\frac{1}{\gamma_0 q_0 (1+a_{N}) -m_N (1+b_{N})}
\end{equation}
where $a_{N}$ and $b_{N}$ are $T$ dependent corrections.
Multiplying the numerator and the denominator in the above formula by
$\gamma_0 q_0 (1+a_{N}) +m_N (1+b_{N})$
we identify the thermal shift of the nucleon pole  as

\begin{equation}
\delta m_N (T)= m_N (b_{N} (T)-a_{N} (T)),
\label{shift}
\end{equation}
While $\delta m_N (T)$ here is of order $T^4$, the
functions $a_{N}(T)$ and $b_{N}(T)$ contain $T^2$ terms\cite{ls},
$a_{N} (T)=g_{A}^{2}\xi +O(T^4)$,
$b_{N} (T)=g_{A}^{2}\xi +O(T^4)$, where $g_A =1.27$ is the usual
axial coupling of the nucleon. They are due to the contribution
of the $\pi N$ self-energy graph at $T\neq 0$.
On the other hand, Eq.(\ref{gamma}) and
(\ref{1}), which are exact, do not contain any terms
involving $g_{A}^{2}$.
This seeming paradox was resolved in Ref.\cite{koi}
in which it was shown that the $g_{A}^{2}$ terms indeed cancel
provided the $\pi N\to N$ scattering is accounted for in the
spectral density of the correlator at $T\neq 0$.
Taking this into consideration we obtain the nucleon contributions to
$C_1$ and $C_2$

\begin{eqnarray}
C_{1} (Q^2,T) &=& \lambda_{1}^{2}(0)
\frac{1-\bar{a}_{N} (T)}{Q^2 +m^{2}_N +\delta m^{2}_N (T)} \nonumber\\
C_{2} (Q^2,T) &=& \lambda_{2}^{2}(0)
\left(1-2\xi-\frac{2}{3}\xi^{2}\right)
\frac{m_{N} (1-\bar{a}_{N}(T))+
\delta m_N (T)}{Q^2 +m^{2}_N +\delta m^{2}_N (T)}
\label{c12}
\end{eqnarray}
where $\bar{a}_{N}(T)=O(T^4)$ is the nucleon wave-function
renormalization due to $\pi X$ self-energy corrections
to the nucleon propagator at $T\neq 0$, where $X\neq N$.
To relate $\delta m_{N}(T)$ and
$\bar{a}_{N} (T)$ to the $T^4$ corrections obtained before,
we match Eq.(\ref{c12}) against
Eq.(\ref{gamma}) and (\ref{1}) keeping terms linear in
$\delta m_N$ and $\bar{a}_{N}$.
To suppress the contribution of higher states, the Borel transformation

\begin{equation}
\hat{B} C(Q^2) = \lim_{\stackrel{Q^2,n\to\infty}{Q^2/n=M^2=const}}
\frac{1}{(n-1)!}(Q^2)^n \left( -\frac{d}{dQ^2}\right)^n C(Q^2)
\label{borel}
\end{equation}
is usually performed in QCD sum rules. Though we neglect the changes
in the continuum induced by temperature, the Borel transformation
is still necessary to kill the subtraction terms proportional to
$T^4$ in Eq.(\ref{gamma}). Applying it, we get

\begin{eqnarray}
\frac{\delta m_N}{m_N}
 &=& -\frac{4M_{2}M^2 e^{m^{2}_{N}/M^2}}{15\lambda^{2}_{N}}
\left( 1+\frac{3\pi^{2}\langle\bar{q}q\rangle }{2M^2 m_N}\right) T^4
\nonumber\\
\bar{a}_{N} &=& -\frac{4M_{2}M^2 e^{m^{2}_{N}/M^2}}{15\lambda^{2}_{N}}
\left( 1-\frac{2m^{2}_N}{M^2}-
\frac{3\pi^{2}m_{N}\langle\bar{q}q\rangle }{M^4}\right) T^4
\label{ma}
\end{eqnarray}
We see that the mass shift is negative.
If $M^2$ varies between $0.7$ and $2$ MeV$^2$
(the stability range of the
$T=0$ sum rules for the nucleon), $\delta m_N$ changes by $30\%$,
while $\bar{a}$ is less stable.
Taking $M=m_N$, $M_2 =0.12$
\footnote{See \cite{va} for a discussion of existing
estimates of $M_2$.}
and $\lambda_{N}^{2}=1.2\cdot 10^{-3}$\, GeV$^{6}$\cite{bl},
we find that the mass shift is rather small:
$\delta m_{N}\approx -2$\, MeV at $T=80$\, MeV. We note that up
till this temperature our estimate for $\delta m_N$ agrees with the
result of Ref.\cite{ls}, obtained using experimental
information on the $\pi N$ forward scattering amplitude.

Taking into account higher twist $s=2$ condensates,
when they become available, one could make the above
predictions more accurate. (Note, however, that $\lambda_{N}^{2}$
is known with an accuracy of 50\%\cite{bl}).
Also, to go to higher $T$ in
this approach it is necessary to known higher spin condensates
which contribute to next terms in the low $T$ expansion of
$\delta m_{N}$.

We have also considered the case of
$\Delta$-isobar. In that case the current in question is\cite{bl}

\begin{equation}
\eta_{\mu} =(u^a C\gamma_{\mu}u^b )u^c\,\epsilon^{abc}
\label{eta1}
\end{equation}
The corresponding correlator $C_{\mu\nu}^{\Delta}(q)$
has a number of tensor structures
already at $T=0$
and is contributed not only by
$\Delta$, but also by the resonance with $J^P =\frac{1}{2} ^{-}$.
However, the structures $g_{\mu\nu}\hat{q}$ and $g_{\mu\nu}$
correspond only to $J^P =\frac{3}{2}$ states. At $T\neq 0$
independent structures proliferate and we again simplify
the situation by putting ${\bf q}=0$.
The calculation is quite similar to the case of nucleon.
The pion matrix elements which give rise to the $T^4/Q^4$
corrections proportional to $g_{\mu\nu}$ are

\begin{equation}
-i\frac{24M_2}{\pi^{4}x^{8}}
((p_{\mu}x_{\nu}+p_{\nu}x_{\mu})\hat{x}(px)-g_{\mu\nu}\hat{x}(px)^2)
\label{dg}
\end{equation}
for the chirality conserving and

\begin{equation}
\frac{4M_2\langle\bar{q}q\rangle}{\pi^2 x^4}
((p_{\mu}x_{\nu}+p_{\nu}x_{\mu})(px)-g_{\mu\nu}(px)^2)
\label{d1}
\end{equation}
for the chirality violating structure.
Integrating over the thermal pion phase space and
going to the momentum representation we get

\begin{equation}
\frac{2M_2}{45}T^4 {\rm ln}Q^2\;\cdot g_{\mu\nu}q_0\gamma_0
\label{dgq}
\end{equation}
and

\begin{equation}
\frac{8M_2\langle\bar{q}q\rangle}{45}\frac{T^4}{Q^2}\;\cdot g_{\mu\nu}
\label{d1q}
\end{equation}
Matching these corrections against the contribution of
$\Delta$ to the spectral density, we get after the Borel
transformation

\begin{eqnarray}
\frac{\delta m_{\Delta}}{m_{\Delta}}
 &=& -\frac{2M_{2}M^2 e^{m^{2}_{\Delta}/M^2}}{45\lambda^{2}_{\Delta}}
\left( 1+\frac{4\pi^{2}\langle\bar{q}q\rangle }
{M^2 m_{\Delta}}\right) T^4
\nonumber\\
\bar{a}_{\Delta} &=& -\frac{2M_{2}M^2 e^{m^{2}_{\Delta}/M^2}}
{15\lambda^{2}_{\Delta}}
\left(1-\frac{2m^{2}_{\Delta}}{M^2}-
\frac{8\pi^{2}m_{\Delta}\langle\bar{q}q\rangle }{M^4}\right) T^4
\label{mad}
\end{eqnarray}
The mass shift is again negative.
Substituting $\lambda_{\Delta}^{2}=2.5\cdot 10^{-3}$ GeV$^{6}$\cite{bl}
and putting $M=m_{\Delta}$, we see that mass shift is much smaller
than the nucleon one:
$\delta m_{\Delta}\approx -0.25$ MeV at $T=80$ MeV.

To summarize, we have used the only presently available non-scalar
condensate of spin 2, the energy-momentum tensor, to calculate
full order $T^4$ corrections to the correlators with quantum numbers of
nucleon and $\Delta$-isobar. Expressed in terms of physical nucleon and
$\Delta$ contributions, these corrections determine the leading low $T$
behavior of the thermal mass shifts, which are of order $T^4$ and
negative, as in the case of $\rho$ and $a_1$ mesons\cite{va}.
The mass shifts are rather small numerically, reaching a few MeV at
$T\sim 100$ MeV. This is an order of magnitude less than what one
would get from the naive scaling,
$m_{N(\Delta)} (T)\sim (-\langle\bar{q}q\rangle _{T})^{1/3}$, motivated
by the results of Ref.\cite{bl}.

\vspace{0.75cm}

I am grateful to the staff of Centre de Physique Th\'eorique
at Luminy (Marseille)
where this work has been done for the warm hospitality extended to me
and, especially, to Chris Korthals Altes for useful discussions.
Special thanks are due to Jay Watson for his help with LATEX figures.
This work was supported in part by the ISF grant M9H000.

\end{document}